\documentclass[aps,prl,reprint,superscriptaddress]{revtex4-1}
\usepackage{hyperref}
\usepackage{braket}
\usepackage{graphicx}
\usepackage{amsmath}
\usepackage[caption=false]{subfig}
\graphicspath{{figures/}}
\usepackage{xr}

\begin{document}
\title{Dirac particle dynamics of a superconducting circuit}
\author{Elisha Svetitsky}
\affiliation{Racah Institute of Physics, the Hebrew University of Jerusalem, Jerusalem, 91904 Israel}
\author{Nadav Katz}
\affiliation{Racah Institute of Physics, the Hebrew University of Jerusalem, Jerusalem, 91904 Israel}

\begin{abstract}     

The core concept of quantum simulation is the mapping of an inaccessible quantum system onto a controllable one by identifying analogous dynamics. We map the Dirac equation of relativistic quantum mechanics in 3+1 dimensions onto a multi-level superconducting Josephson circuit. Resonant drives determine the particle mass and momentum and the quantum state represents the internal spinor dynamics, which are cast in the language of multi-level quantum optics. The degeneracy of the Dirac spectrum corresponds to a degeneracy of bright/dark states within the system and particle spin and helicity are employed to interpret the multi-level dynamics. We simulate the Schwinger mechanism of electron-positron pair production by introducing an analogous electric field as a doubly degenerate Landau-Zener problem. All proposed measurements can be performed well within typical decoherence times. This work opens a new avenue for experimental study of the Dirac equation and provides a tool for control of complex dynamics in multi-level systems.

\end{abstract}

\maketitle

Paul Dirac's celebrated equation for the electron  \cite{Dirac1928} combines special relativity and quantum mechanics, predicting a host of phenomena such as antimatter and the spin $\frac{1}{2}$ degree of freedom. In the Dirac equation a single particle is described by a spinor comprising four functions of space and time. In contrast, quantum information and simulation \cite{Feynman1982,RevModPhys.86.153,Cirac2012,Buluta108} experiments are invariably performed in systems obeying the non-relativistic Schrodinger equation. In order to experimentally explore the Dirac equation in the laboratory, an appropriate mapping is required to enable a controllable quantum system to mimic a relativistic one. Here, we study the Dirac equation by mapping the internal spinor dynamics to a four-level system, resulting in novel intuition into complex multi-level dynamics previously studied in the context of quantum optics.  While other works have made strides towards scalable simulations of quantum field theories \cite{0034-4885-79-1-014401,Martinez2016,Alba201364,jordan2012quantum}, we focus on the first-quantized Dirac equation to provide new perspective on coherent control of multi-level systems, promising ingredients for resource-efficient quantum computation \cite{lanyon2009simplifying,fedorov2012implementation} but often overlooked due to the complexity of their dynamics. In principle our scheme can be realized in a variety of systems with the appropriate level structure. We propose the use of superconducting Josephson circuits \cite{Clarke2008,Devoret2013} and show how to engineer the necessary level structure, motivated by their uniqueness in combining long coherence times, flexibility of design, and straightforward control.

The Dirac equation can be written in Hamiltonian form, 
$i\hbar\frac{\partial\psi}{\partial t}=H\psi$ \cite{Dirac1928}. In 1+1 and 2+1 dimensions $\psi$ is a two-component spinor with  positive and negative energy components. In 3+1 dimensions the spinor has four components, reflecting the additional property of spin $\frac{1}{2}$. Spinor components are functions of space and time coupled together by the Hamiltonian. Our goal is to realize this Hamiltonian with a controllable quantum system.

\begin{figure}
	\includegraphics[trim=1cm 0 1cm 0, width=0.67\linewidth]{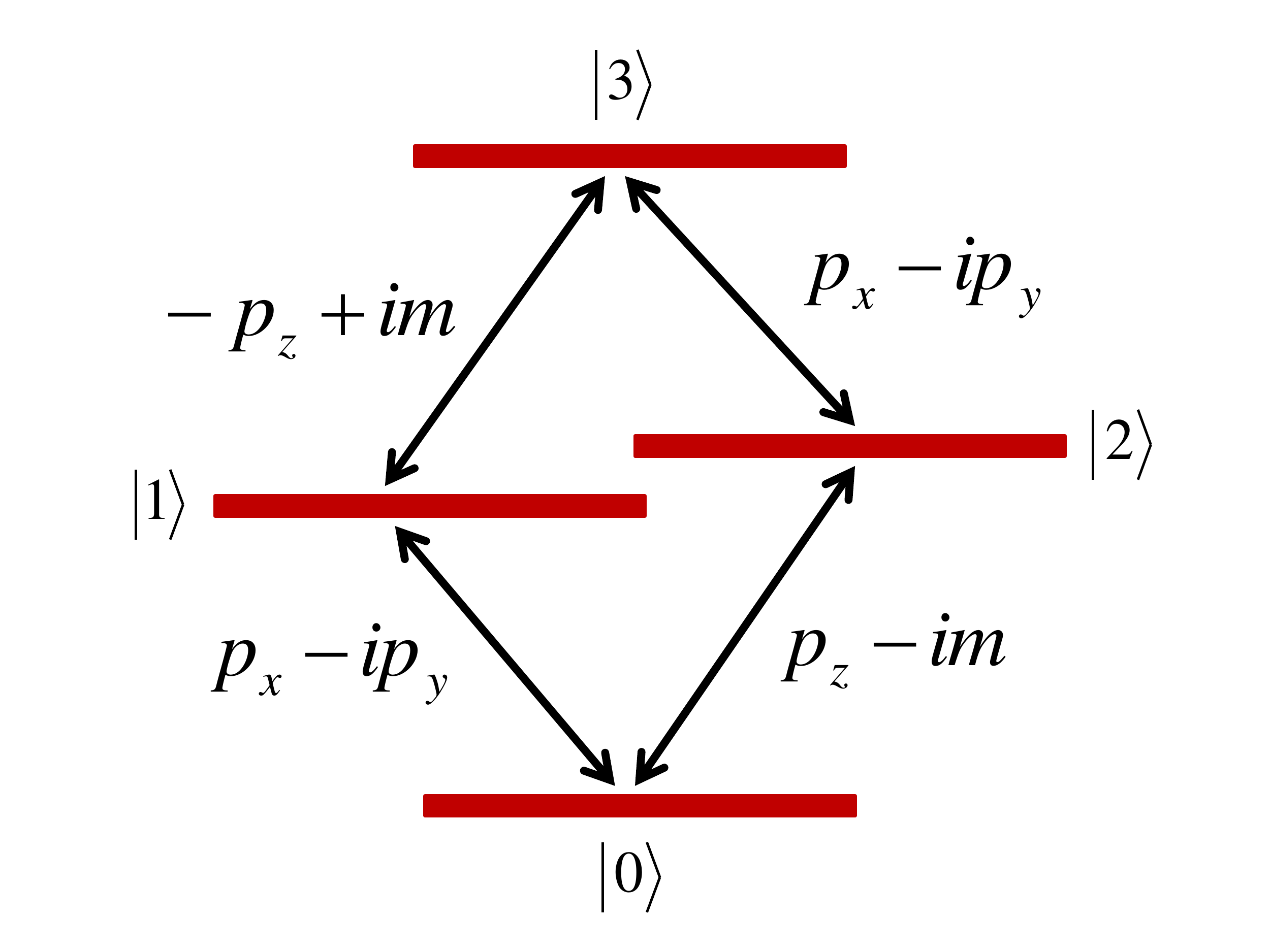} 
	\caption{Diamond level structure with resonant nearest-neighbor couplings. The four drive parameters correspond to particle mass and momentum, and the four level state is mapped to the spinor components of a Dirac particle.}
	\label{fig:Fig1}
\end{figure}

\begin{figure}
	\includegraphics[trim=0 2cm 0 0, width=1\linewidth]{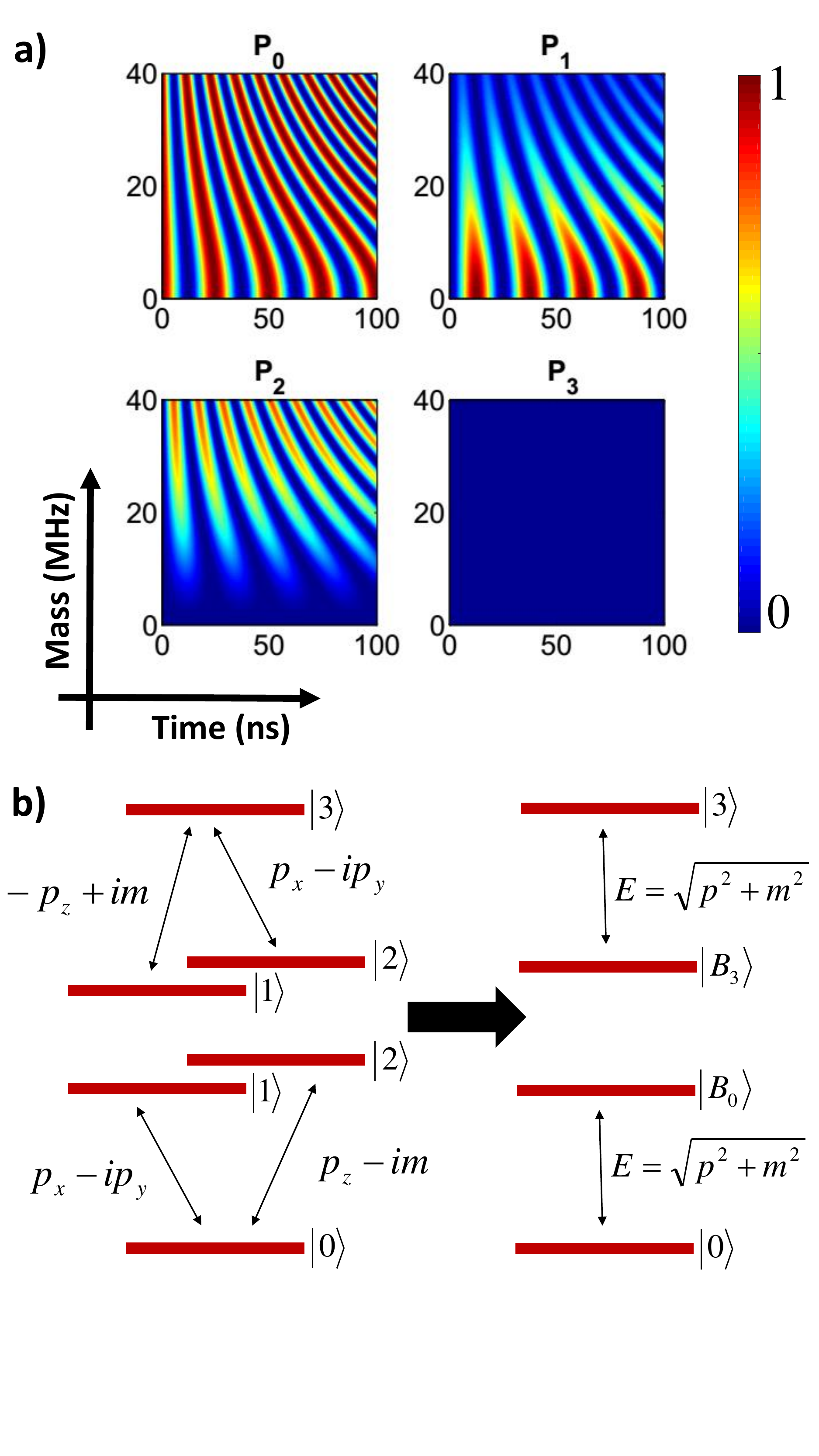} 
	\caption{Simulation of the free Dirac equation. (a) Dynamics of a Dirac particle with momentum $\frac{\vec{p}}{2\pi}=\left(20\:\textrm{MHz},\:0,\:0\right)$  and a range of mass values. The population oscillates with a single frequency between the ground state and a superposition of states $\ket{1},\ket{2}$, never populating $\ket{3}$, a result of the degeneracy of the Dirac spectrum. (b) Quantum optics interpretation: the diamond level configuration can be separated into overlapping $\Lambda$ and V systems, where for Dirac drives the bright state $\ket{B_0}$ of the V system coincides with the dark state of the $\Lambda$ system orthogonal to its bright state $\ket{B_3}$.}
	\label{fig:Fig2}
\end{figure}

We begin by considering a four-level system with nearest-neighbor transitions in a diamond configuration. The transitions are resonantly driven with amplitudes as shown in Fig. 1. Note that with this level structure the complex phases of the drives cannot all be absorbed into the bare levels as for a ladder configuration. Choosing to order the basis such that $\ket{\psi}\:=\:c_0\ket0+c_1\ket1+c_2\ket2+c_3\ket3=(c_0 \: c_3 \: c_2 \: c_1 )^{T}$ we obtain the Hamiltonian

\begin{equation}
H=\begin{pmatrix}
0 & 0 & p_{z}-im & p_{x}-ip_{y}\\
0 & 0 & p_{x}+ip_{y} & -p_{z}-im\\
p_{z}+im & p_{x}-ip_{y} & 0 & 0\\
p_{x}+ip_{y} & -p_{z}+im & 0 & 0
\end{pmatrix}\label{eq:1}
\end{equation}

This is the Dirac Hamiltonian in the supersymmetric representation \cite{thaller2013dirac} for a plane wave with mass $m$ and momentum  $\vec{p}=\left(p_{x},\:p_{y},\:p_{z}\right)$ (with natural units $\hbar=c=1$). For such a state the four spinor components have the same spatial dependence, $\psi\sim exp(i\vec{p}\cdot\vec{r})$, allowing a reduction from a continuous to a discrete basis by suppressing the common spatial factor. This mapping is limited to a momentum eigenstate, i.e. a single Fourier component; wave packets can be constructed in post-processing of ensemble measurements. This representation was first introduced to quantum simulation by \cite{Lamata2007} in the context of trapped ions and experimentally demonstrated by \cite{Gerritsma2010} in 1+1 dimensions as a spatially dependent Rabi system. In this work we focus on superconducting circuits which do not have a motional degree of freedom and therefore require a different mapping and interpretation of particle momentum, while maintaining validity of insights provided by the analogous Dirac dynamics. Moreover, our proposal brings the Dirac equation in 3+1 dimensions well within current experimental capabilities.

We realize the necessary diamond level structure with a pair of nominally identical, capacitively coupled superconducting transmon qubits \cite{houck2009life,Barends2013}. In identifying suitable circuit parameters for a multi-level experiment we seek to avoid crosstalk between different transitions while respecting the inherent tradeoff between Josephson qubit anharmonicity and sensitivity to charge noise \cite{Koch2007,PhysRevB.77.180502}. As shown in  \cite{SuppNote}, suitable parameters are a bare first transition frequency of 5 GHz, anharmonicity of 300 MHz, and qubit-qubit coupling of 100 MHz, all readily achievable with standard fabrication techniques. The coupling hybridizes the $\ket{01},\ket{10}$ levels and results in a splitting of 200 MHz. Similar hybridization and level repulsion within the $\left\{\ket{20},\ket{11},\ket{02}\right\} $ manifold causes the highest-lying state to be shifted upwards by 100 MHz, resulting in four distinct transitions which can be addressed individually. Thus, all transitions are separated by at least 100 MHz, setting a reasonable limit on bandwidth and drive amplitudes. Multi-tone signals with phase control of each component can be synthesized with commercial electronics, making simultaneous resonant driving of several transitions straightforward. Throughout this work we neglect decoherence; typical coherence times for transmon qubits are an order of magnitude longer than required for our purpose.

Figure 2a shows the dynamics of a Dirac particle with $\vec{p}=\left(20\:\textrm{MHz},\:0,\:0\right)$ and a range of mass values. The populations oscillate smoothly with a single frequency between the ground state and a superposition of the $\ket1,\ket2$ states, never populating the  $\ket3$ state. This result is initially surprising considering the usually complicated nature of multi-level dynamics, but can be explained with the insight of bright and dark states from quantum optics. To do so we split the four-level system into two three-level systems, one with a V configuration and the other with a $\Lambda$ configuration, with the middle states common to both (Fig. 2b). For a V system initially in the ground state the population can be shown to oscillate between the $\ket0$ state and a superposition 'bright state' (denoted $\ket{B_{0}}$) of the $\ket1,\ket2$ levels with a single frequency given by the pythagorean sum of the coupling amplitudes, in this case the relativistic energy $E=\sqrt{|p|^{2}+m^{2}}$. The superposition state in the $\left\{ \ket1,\ket2\right\} $ subspace orthogonal to the bright state is never populated and is termed a 'dark state'. The same can be derived for the $\Lambda$ system, whose bright state we denote $\ket{B_3}$. Remarkably, for Dirac couplings the bright (dark) state of the V system coincides with the dark (bright) state of the $\Lambda$ system and the transitions to the highest state thus interfere destructively, preventing population of the $\ket3$ state. These simplified dynamics can be appreciated in light of the Dirac equation by taking the diagonalized Dirac Hamiltonian and factoring it into single-qubit operators, $diag(E,E,-E,-E)=E\cdot I\otimes\sigma_z$. In addition to revealing the single frequency $E$, this form shows how, despite appearances, the degeneracy limits dynamics to a two-dimensional Hilbert space.

\begin{figure}
	\includegraphics[trim=7cm 0 9cm 0, width=1\linewidth]{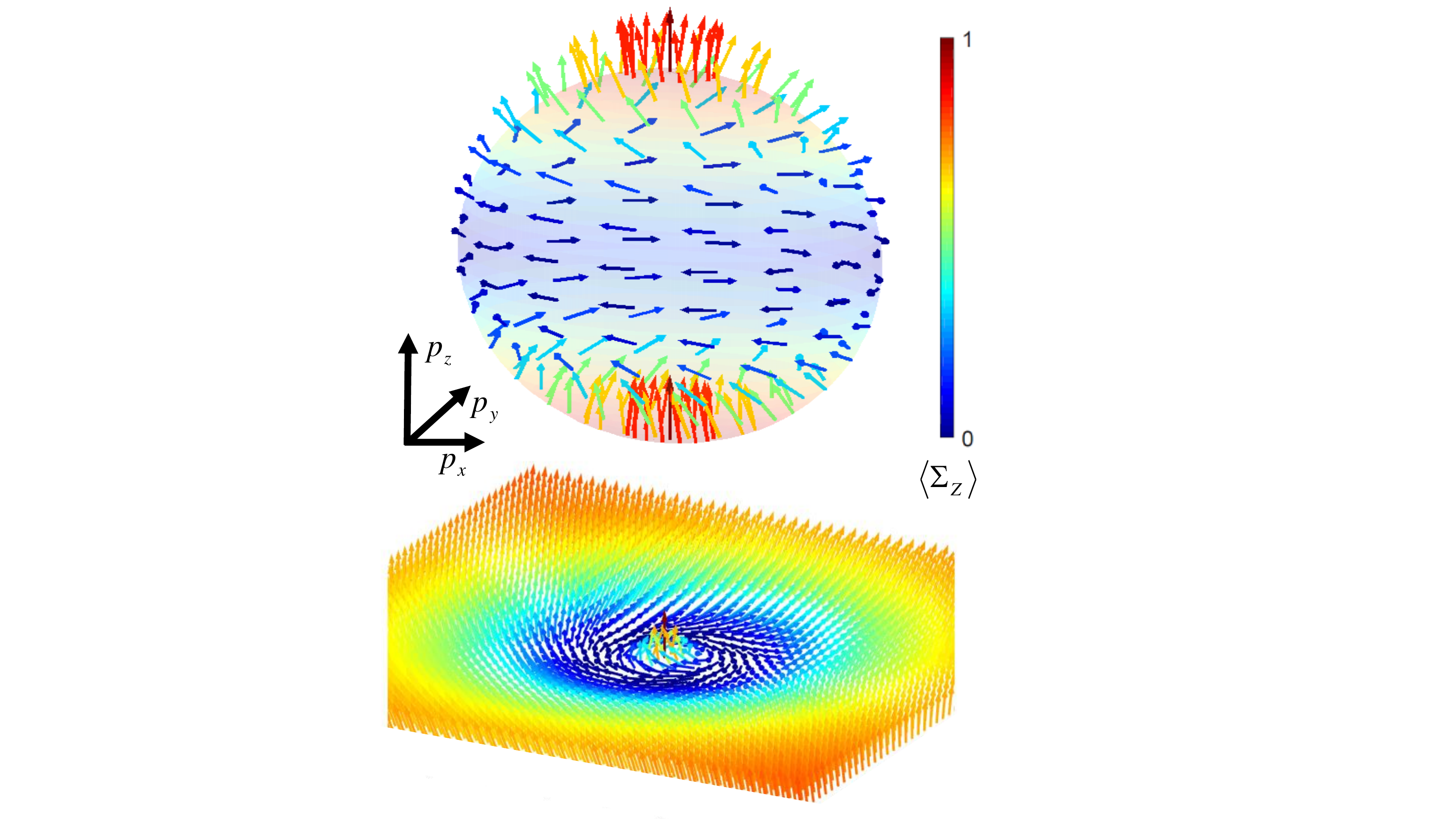} 
	\caption{Particle spin, helicity conservation and the bright state. The Bloch vector of the bright state corresponds to particle spin, here shown in momentum space on a sphere of constant energy  $\frac{|p|}{2\pi}=\frac{m}{2\pi}=20$  MHz. At $t=0$ the spin points up and the radial component equals the helicity. Along the equator helicity is zero and the bright state is tangent to the sphere. For nonzero $p_z$ the particle has finite helicity and the spin obtains a radial component.}
	\label{fig:Fig3}
\end{figure}

Previous work \cite{Lamata2007,Gerritsma2010,Roos2011,lee2015tachyon} on simulating the Dirac equation with trapped ions was limited by experimental considerations to the 1+1 dimensional case, forfeiting the spin $\frac{1}{2}$ property which was the original impetus behind Dirac's work \cite{Dirac1928}. Our system enables simulation in 3+1 dimensions, preserving the spin degree of freedom and providing another handle on multi-level dynamics, specifically the details of the bright state. The components of the spin operator are $\Sigma_i\equiv \frac{1}{2}\begin{pmatrix}
\sigma_i & 0\\
0 & \sigma_i \\ \end{pmatrix}$ ($i=x,y,z$). With our basis ordering $\langle \vec{\Sigma} \rangle$ is the sum of two unnormalized Bloch vectors, one for each of the $\left\{\ket{0},\ket{3}\right\}$ and $\left\{\ket{2},\ket{1}\right\}$ manifolds, coupled together by the Hamiltonian. In relativistic quantum mechanics particle spin is only part of the total angular momentum and is not conserved separately. To obtain a conserved quantity in our system we turn to the helicity, defined as the spin projection along the momentum axis, $\hat{h}=\frac{\vec{p}\cdot\vec{\Sigma}}{|p|}$.

Figure 3 shows the bright state as a function of momentum on a sphere of constant energy, as well as a stereographic projection where the south pole is mapped to the origin and the north pole to infinity. Since the spin points up at $t=0$ when the system is in the ground state, along the equator where $p_z=0$ the helicity is zero and the spin is orthogonal to the momentum at all times. Adding a $p_z$ component endows the particle with non-zero helicity and the bright state obtains a radial component. At the poles the spin is locked pointing up as required by helicity conservation. Experimentally, the momentum direction can be chosen by setting the phase of the drives on the $0-1$ and $2-3$ transitions, and the spin can be detected experimentally with two-qubit tomography \cite{Steffen1423} of the $\left\{\ket{1},\ket{2}\right\}$ subspace. We note that the topology of the spin texture is trivial, and that a topologically non-trivial spectrum can be obtained with a 'modified' Dirac equation to mimic topological insulators \cite{shen2013topological}.

\begin{figure*}
	\includegraphics[trim=3cm 1cm 8cm 2cm,width=1\linewidth]{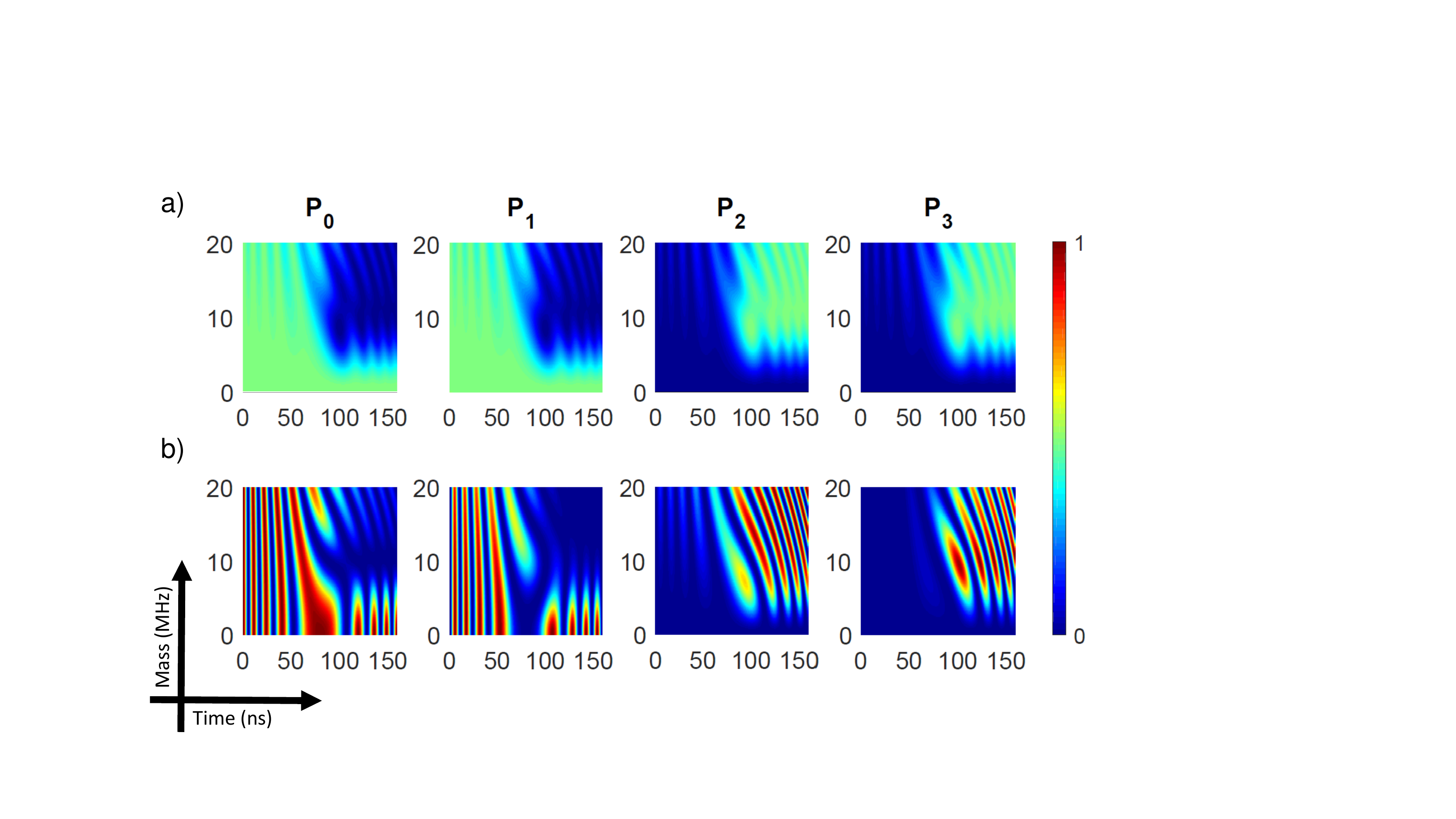} 
	\caption{Dirac particle in an electric field. (a) Populations over time for initial state $\ket{\pm}_{01}=\frac{\ket{0}\pm\ket{1}}{\sqrt{2}}$ and momentum chirped at rate $(10\: \textrm{MHz})^2$ to simulate an electric field. Population is adiabatically transferred to $\ket{\mp}_{23}=\frac{\ket{2}\mp\ket{3}}{\sqrt{2}}$ at large mass and remains in the initial state for low mass. (b) Same as (a) but for initial state $\ket0$. At large mass the ground state population is transferred to the $\left\{\ket{2},\ket{3}\right\}$ manifold with beating between states $\ket{\pm}_{23}$ and $\ket{\mp}_{23}$. For small mass the population remains in the  $\left\{\ket{0},\ket{1}\right\}$ manifold and beats between states $\ket{\pm}_{01}$ and $\ket{\mp}_{10}$.}
	\label{fig:Fig4}
\end{figure*}

\begin{figure}
	\includegraphics[trim=5cm 0 5cm 0, width=1\linewidth]{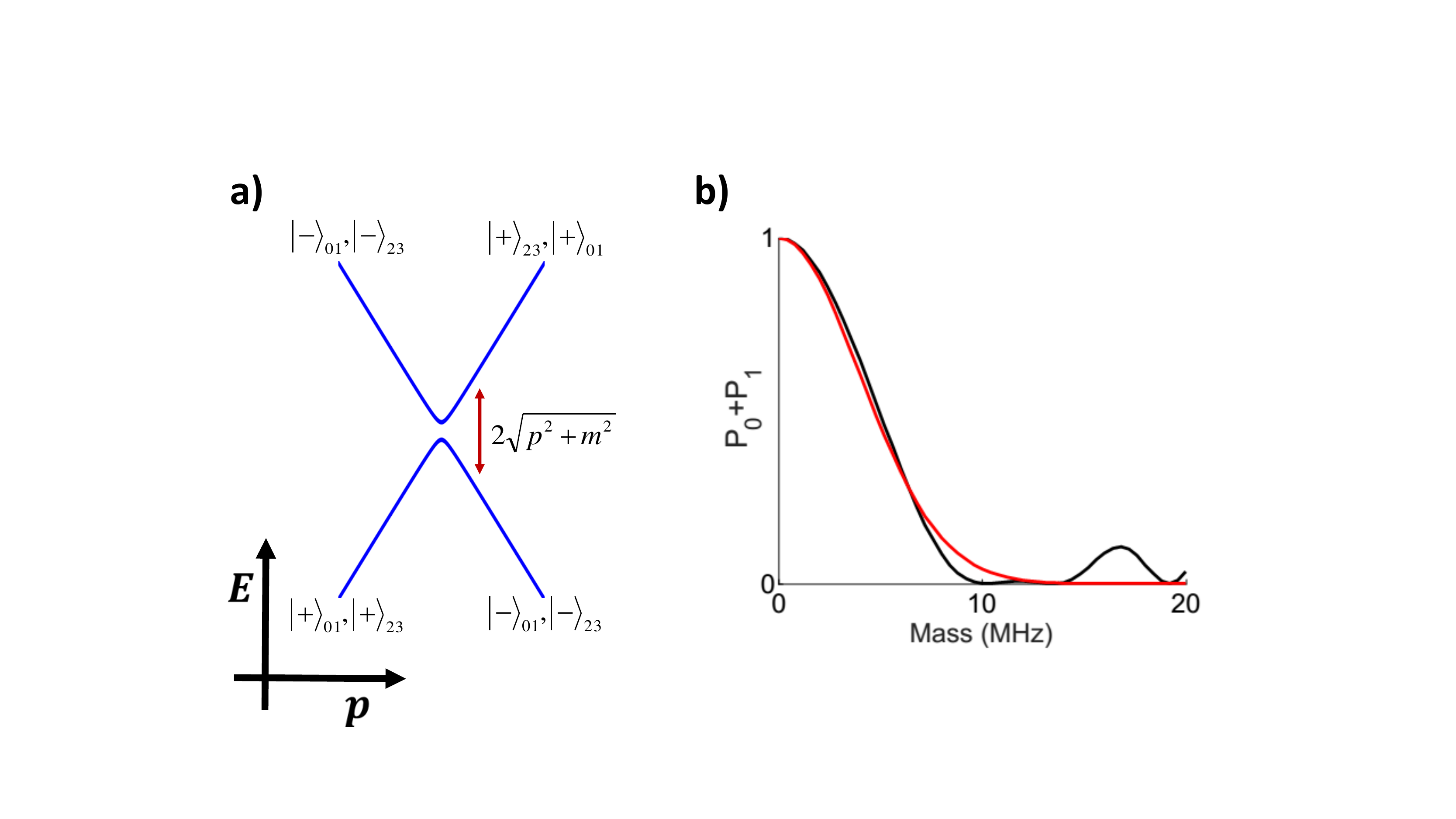} 
	\caption{Equivalence of a degenerate Landau-Zener problem and the Schwinger mechanism. (a) As shown in the previous figure, the ground state $\ket0$ contains two adiabatic  components, each of which is transferred to its counterpart in the $\left\{\ket{2},\ket{3}\right\}$ manifold. Degenerate states are not coupled. The level separation reaches a minimum of $2m$, defining the maximal chirp rate for adiabatic transfer. (b) Final $\left\{\ket{0},\ket{1}\right\}$ subspace population for a range of mass values (black) and the Landau-Zener tunneling or Schwinger pair production formula (red).  The deviations at large mass are due to the finite momentum limits relative to the gap, a constraint of the system's anharmonicity.} \label{fig:Fig5}
\end{figure}

Returning to the bright/dark state analysis, with such a drive-dependant interference effect it is natural to inquire into the effect of time-dependant drive amplitudes. These simulate electric fields in the gauge $\vec{\epsilon}=-\frac{\partial\vec{A}}{\partial t}$ (i.e. no scalar potential; here $\vec{\epsilon}$ is a simulated electric field and $\vec{A}$ is the vector potential) by shifting the momentum $\vec{p}\rightarrow\vec{p}-e\vec{A}$. A spatially inhomogeneous vector potential is not diagonal in the momentum representation, ruling out a straightforward simulation of a magnetic field. The Dirac equation in the presence of an electric field gives rise to electron-positron pair production in a 'dielectric breakdown of the vacuum', worked out by Sauter \cite{Sauter1931} for the first-quantized Dirac equation and by Schwinger \cite{Schwinger1951}, for whom the effect is usually named, with quantum electrodynamics. This effect is simulated by ramping the drive amplitude $p_{x}$ at a rate $\epsilon_{x}=(10\:\textrm{MHz})^{2}$ from an initial value $p_{x}=-50$ MHz to a final value  $p_{x}=50$ MHz. Beginning in the state $\ket\pm_{01}=\frac{1}{\sqrt{2}}\left(\ket0\pm\ket1\right)$ (an eigenstate for  $m\ll\left|p_{x}\right|$), population is transferred to the state $\ket\mp_{23}=\frac{1}{\sqrt{2}}\left(\ket2\mp\ket3\right)$ (Fig. 4a). For a system initially in the ground state $\ket0=\frac{1}{\sqrt{2}}\left(\ket{+}_{01} +\ket{-}_{01}\right)$, interference between the transitions leads to oscillations in the $\left\{ \ket2,\ket3\right\}$ manifold (Fig. 4b).

The Schwinger formula for the pair production probability is $P=exp(-\pi\frac{m^{2}}{\epsilon})$, where $m$ is the particle mass and $\epsilon$ the electric field magnitude. The Schwinger limit $\epsilon_{s}=m^{2}$ for observing the effect is yet inaccessible even for the most powerful lasers \cite{1742-6596-672-1-012020} due to the size of the electron mass, but in our simulation the mass is a tunable experimental parameter. The similarity of this formula to the two-level Landau-Zener tunneling probability has already been noted for two-component spinors \cite{rau1996reversible,PhysRevD.78.096009}. Since $\ket\pm_{01}$ do not couple to each other but only to $\ket\mp_{23}$, respectively, we can understand the multi-level dynamics of the chirped 3+1 dimensional Dirac equation by factoring it into two separate Landau-Zener problems, illustrated in Fig. 5a. For a fast sweep rate (large electric field) compared to the minimum energy gap (particle mass) the non-adiabatic tunneling between branches leaves population in the initial $\left\{\ket0,\ket1\right\}$ manifold. We therefore identify the population remaining in that manifold with pair production which decreases with large mass. In Fig. 5b we show the final population of the $\left\{\ket0,\ket1\right\}$ manifold and compare to the Schwinger/Landau-Zener expression. Gaussian suppression of pair production with particle mass is seen clearly.

In conclusion, the Dirac equation in 3+1 dimensions is mapped to a controllable quantum system with a diamond level structure. We propose a new realization of such a level structure with a superconducting architecture by capacitively coupling two three-level Josephson qubits and identify feasible circuit parameters. The populations of four of the dressed states mimic the internal spinor dynamics of a Dirac particle whose mass and momentum are determined by the drive parameters. Thus we translate concepts between multi-level quantum optics and the Dirac equation. Multi-level interference effects are explained both in terms of quantum optics and the degeneracy of the Dirac spectrum. In contrast to previous works which studied the Dirac equation in lower dimensions, our scheme retains properties such as spin $\frac{1}{2}$ and helicity which are explicitly demonstrated in our system. We incorporate an electric field by chirping the drive amplitudes, simulating Schwinger pair production as a four-level Landau-Zener problem.

Wave packet and multi-mode spatio-temporal effects such as Zitterbewegung \cite{schrodinger1930kraftefreie} are observable in our scheme by superposing different plane wave measurements. Further study may extend this mapping to non-diagonal operators, e.g. spatially inhomogeneous vector potentials or 'synthetic' magnetic fields \cite{ray2014observation,Roushan2017}. Scaling to multi-particle dynamics requires an adaptation of the Jordan-Wigner transformation suitable for the specific multi-level structure in order to maintain fermionic statistics \cite{ortiz2001quantum}. Topological condensed matter systems  \cite{shen2013topological} can be modeled in this context by also explicitly breaking symmetries in our Dirac Hamiltonian in a controlled manner.

We thank Benjamin Svetitsky, Andreas Wallraff, and Enrique Solano for fruitful discussions. This work is supported by the European Research Council Project No. 335933.


%

\cleardoublepage

\widetext
\begin{center}
\textbf{\large Supplementary material: Dirac particle dynamics of a superconducting circuit}

\text{Elisha Svetitsky$^1$ and Nadav Katz$^1$}

\textit{$^1$Racah Institute of Physics, the Hebrew University of Jerusalem, Jerusalem, 91904 Israel}
\end{center}

\setcounter{equation}{0}
\setcounter{figure}{0}
\setcounter{table}{0}
\setcounter{page}{1}
\makeatletter
\renewcommand{\theequation}{S\arabic{equation}}
\renewcommand{\thefigure}{S\arabic{figure}}

	\section*{Diamond level structure with superconducting circuits}
	
	Here we show how the level structure illustrated in Fig. 1 of the main text is obtained with superconducting qubits. The requisite properties are a diamond level structure with nearest-neighbor transitions and a unique frequency for each transition.
	
	Superconducting qubits are anharmonic LC oscillators with unequally spaced levels in a ladder configuration. They can be designed for a chosen anharmonicity (defined here as $\omega_{12}-\omega_{01}$), but large anharmonicity comes at a price of increased sensitivity to charge noise and resultant dephasing. In typical two-level experiments the anharmonicity  imposes a limit on the Rabi frequency, since power broadening induces 'leakage' to higher levels. This problem is compounded for multi-level experiments of the kind described in the main text: the circuit is excited by a multi-tone pulse where each frequency component is intended for a specific transition but is seen by the entire spectrum, and one relies on sufficient detuning from unwanted transitions to prevent crosstalk. 
	
	A simple diamond level structure can be obtained by coupling two two-level systems and identifying the $\big(\ket{00},\ket{01},\ket{10},\ket{11}\big)$ states as $\big(\ket{0},\ket{1},\ket{2},\ket{3}\big)$, but this results in identical frequencies for the $\ket{0}\leftrightarrow\ket{1}$   $\big(\ket{0}\leftrightarrow\ket{2}\big)$ , $\ket{2}\leftrightarrow\ket{3}$  $\big(\ket{1}\leftrightarrow\ket{3}\big)$ transitions, rendering individual addressability impossible. To obtain a fully distinct set of transition frequencies we must consider the levels $\ket{02},\ket{20}$   which are usually justifiably neglected if the single-qubit anharmonicity is sufficiently large relative to both the Rabi frequency and the qubit-qubit coupling. For sufficiently strong coupling the $\ket{11},\ket{02}$, and $\ket{20}$ levels hybridize and repel the $\ket{11}$ upwards, producing the desired frequency shift.

	Writing $\omega_0$ for the first transition, $\kappa$ for the anharmonicity ($\kappa<0$) and $g$ for the qubit-qubit coupling (Fig. S1a), the dressed eigenstates are found by diagonalizing

	\begin{equation}
	H =\begin{pmatrix}
	0 & 0 & 0 & 0 & 0 & 0 \\
	0 & \omega_0 & g & 0 & 0 & 0 \\
	0 & g & \omega_0 & 0 & 0 & 0 \\
	0 & 0 & 0 & 2\omega_0+\kappa & 0 & g\sqrt{2} \\
	0 & 0 & 0 & 0 & 2\omega_0+\kappa & g\sqrt{2} \\
	0 & 0 & 0 & g\sqrt{2} & g\sqrt{2} & 2\omega_0 \\
	
	\end{pmatrix}
	\end{equation}

	For $\kappa=-3g$ we get an effective anharmonicity of $g$ (Fig. S1b). States not part of the desired diamond structure are also detuned by at least this amount from the nearest transition. For transmon qubits, reasonable parameters are $\kappa=-300$ MHz and $g=100$ MHz.

	To confirm that leakage to unwanted states is small we simulate the complete dynamics within a nine dimensional Hilbert space spanned by the first three levels of each qubit. The bare Hamiltonian is written

	\begin{equation}
	H_0 = \omega_{01}(a^\dagger a + b^\dagger b) + \frac{\kappa}{2} (a^\dagger a^\dagger a a + b^\dagger b^\dagger b b) +  g (a^\dagger b + b^\dagger a) 
	\end{equation}

	and the multi-tone excitation field is

	\begin{equation}
	\Omega(t) = V_{01}e^{i\omega_{01}t}+V_{02}e^{i\omega_{02}t}+V_{13}e^{i\omega_{13}t}+V_{23}e^{i\omega_{23}t}
	\end{equation}
	
	where the $V_{ij}$ are Dirac amplitudes. When acting on a single qubit this adds to the Hamiltonian a term
	
	\begin{equation}
	H_{drive} = \Omega(t) a^\dagger + \Omega(t)^*a 
	\end{equation}

	to give the total Hamiltonian $H=H_0+H_{drive}$. The populations in the bare qubit basis are shown in Fig. S2, and Fig. S3 shows the populations of the diamond eigenstates constructed from the previous figure. Depending on the details of the circuit, the diamond eigenstates may be measured directly rather than obtained in post-processing.

	These results are obtained by driving the diamond system through one of its constituent qubits and naively assuming the matrix elements of $H_{drive}$ to be identical for all transitions. Since this is not strictly correct, the results deviate from the Dirac Hamiltonian more than experimentally necessary, since experiments with superconducting qubits routinely involve calibrating drive amplitudes for each transition on-chip. This makes detailed calculations of matrix elements unnecessary and promises better conformity than demonstrated here with the ideal dynamics shown in the main text.

	\section*{Bell basis perspective on Dirac dynamics}

	In previous works \cite{Suchowski2011,Svetitsky2014} it was shown that identifying the states of a four-level system with an entangled two-qubit 'Bell basis' allows the dynamics to be factored into local Bloch sphere rotations. Using the basis

	\begin{equation}
	\begin{array}{cc}
	\ket{0}=\frac{\ket{00}+\ket{11}}{\sqrt{2}} & \ket{1}=\frac{\ket{01}+\ket{10}}{\sqrt{2}} \\
	\\
	\ket{2}=\frac{\ket{01}-\ket{10}}{\sqrt{2}} & \ket{3}=\frac{\ket{00}-\ket{11}}{\sqrt{2}}
	\end{array}
	\end{equation}

	the Dirac Hamiltonian with $\vec{p}=p_{x}\hat{x}$ can be written $H=I\otimes(p_{x}\sigma_{x}+m\sigma_{y})$, describing resonant single-qubit Rabi oscillations with frequency $E=\sqrt{p^{2}+m^{2}}$.
	This representation explains why Dirac couplings result in effective single-qubit dynamics per Fig. 2.
	An electric field causes the Rabi vector of the second qubit to trace a path from the negative to the positive $\hat{x}$ axis, rotating the qubit's plane of precession. For a large electric field the plane of precession changes faster than the Rabi frequency, resulting in rotation around the $z$ axis. As can be seen from the Bell basis, a $z$ rotation on the second qubit directly couples the $\left\{\ket0,\:\ket1\right\}$ and $\left\{\ket2,\:\ket3\right\}$ manifolds, disrupting the adiabatic population transfer. The Bloch vector is most susceptible to this $z$ gate when the Rabi frequency nears its minimum of $m$, illucidating the mass dependence of the Schwinger formula.

	\begin{figure*}
		\includegraphics[width=1\linewidth]{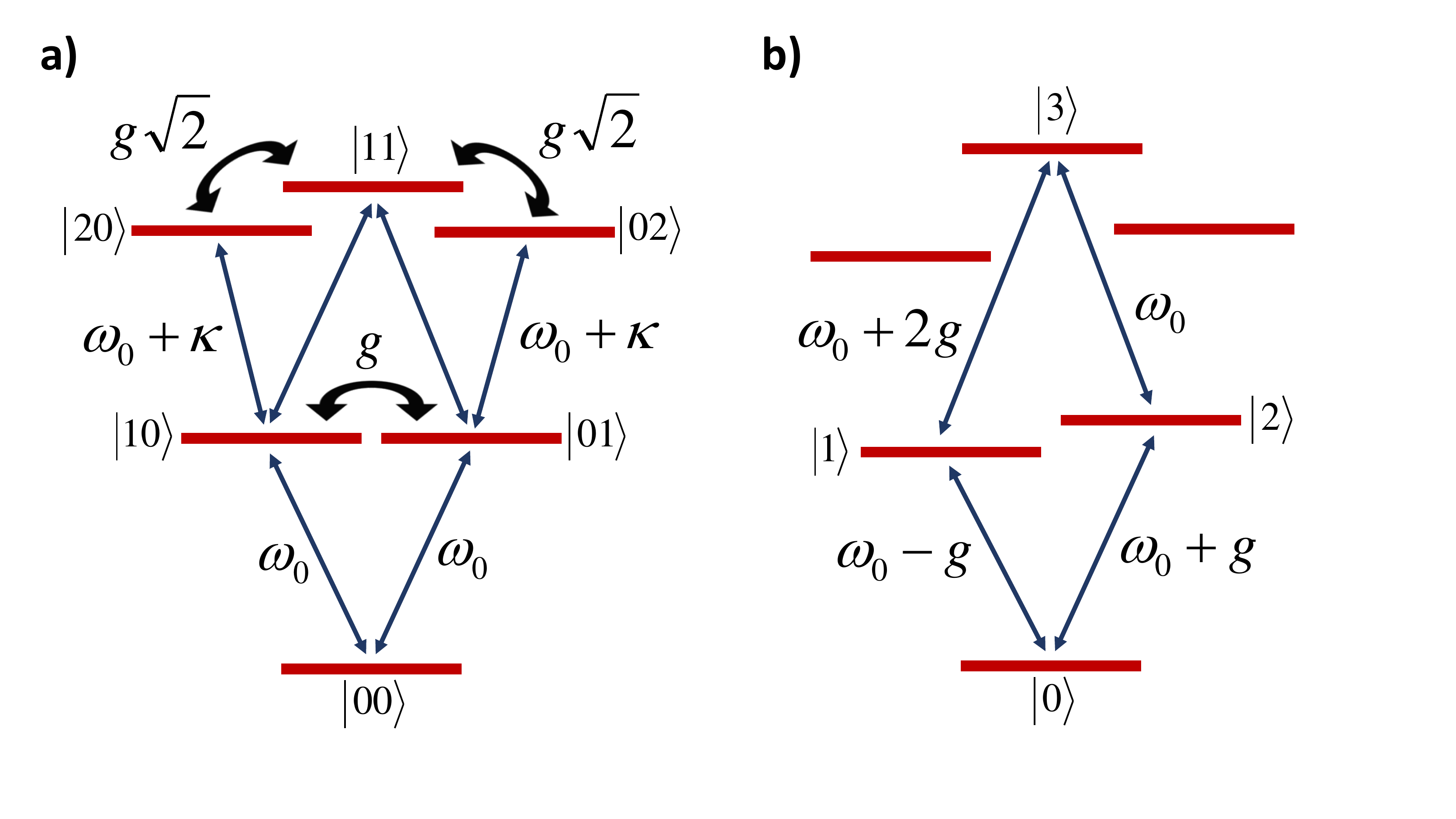} 
		\caption{Level structure of coupled anharmonic oscillators. a) Bare level structure. b) Dressed states for $\kappa=-3g$.}
		\label{fig:S1a}
	\end{figure*}

	\begin{figure*}
		\includegraphics[trim=0cm 0 0 0, width=1\linewidth]{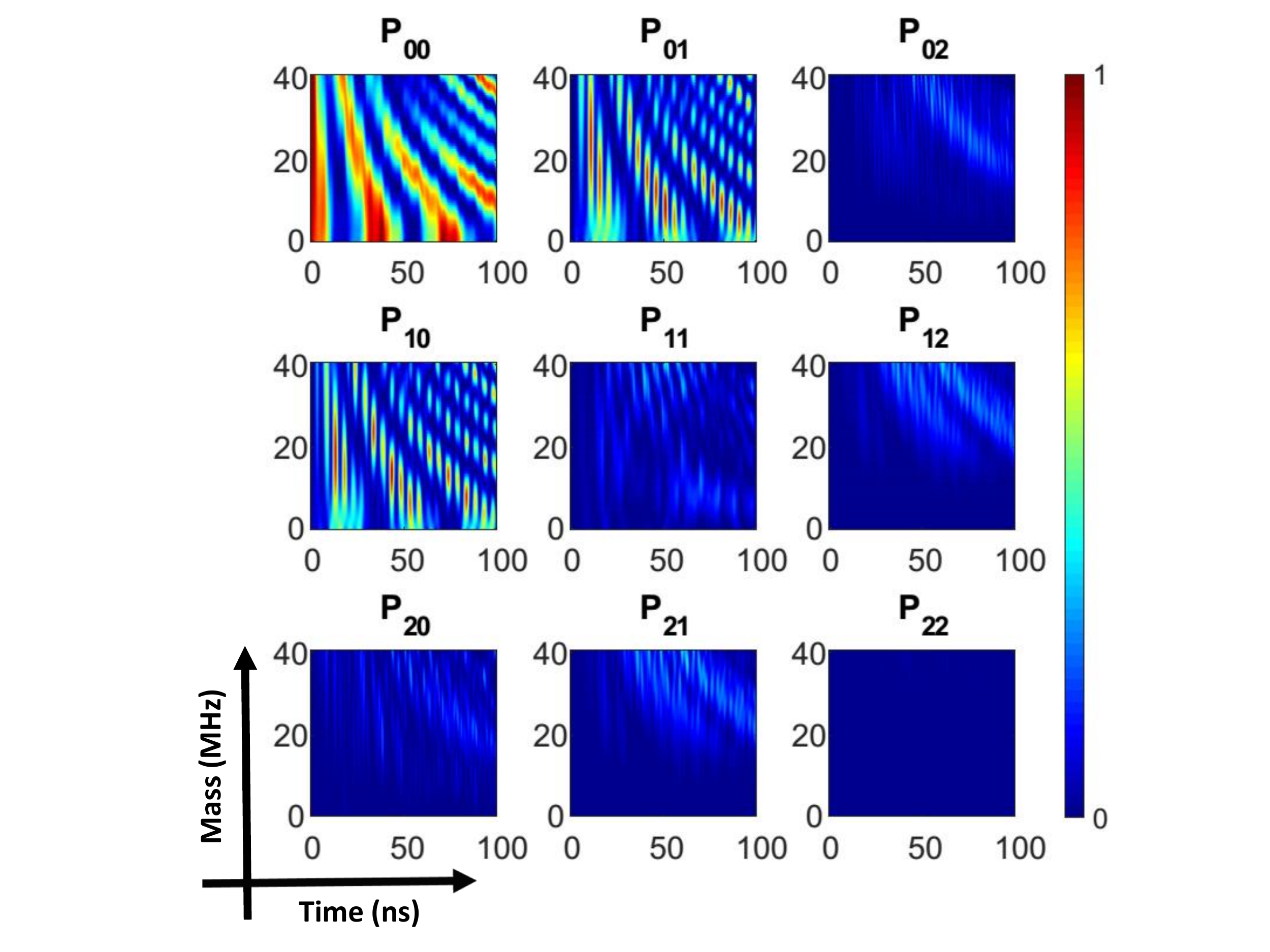} 
		\caption{Populations of bare qubit states. Small leakage to unwanted states occurs at large drive amplitudes due to power broadening.}
		\label{fig:S2}
	\end{figure*}

	\begin{figure*}
		\includegraphics[width=1\linewidth]{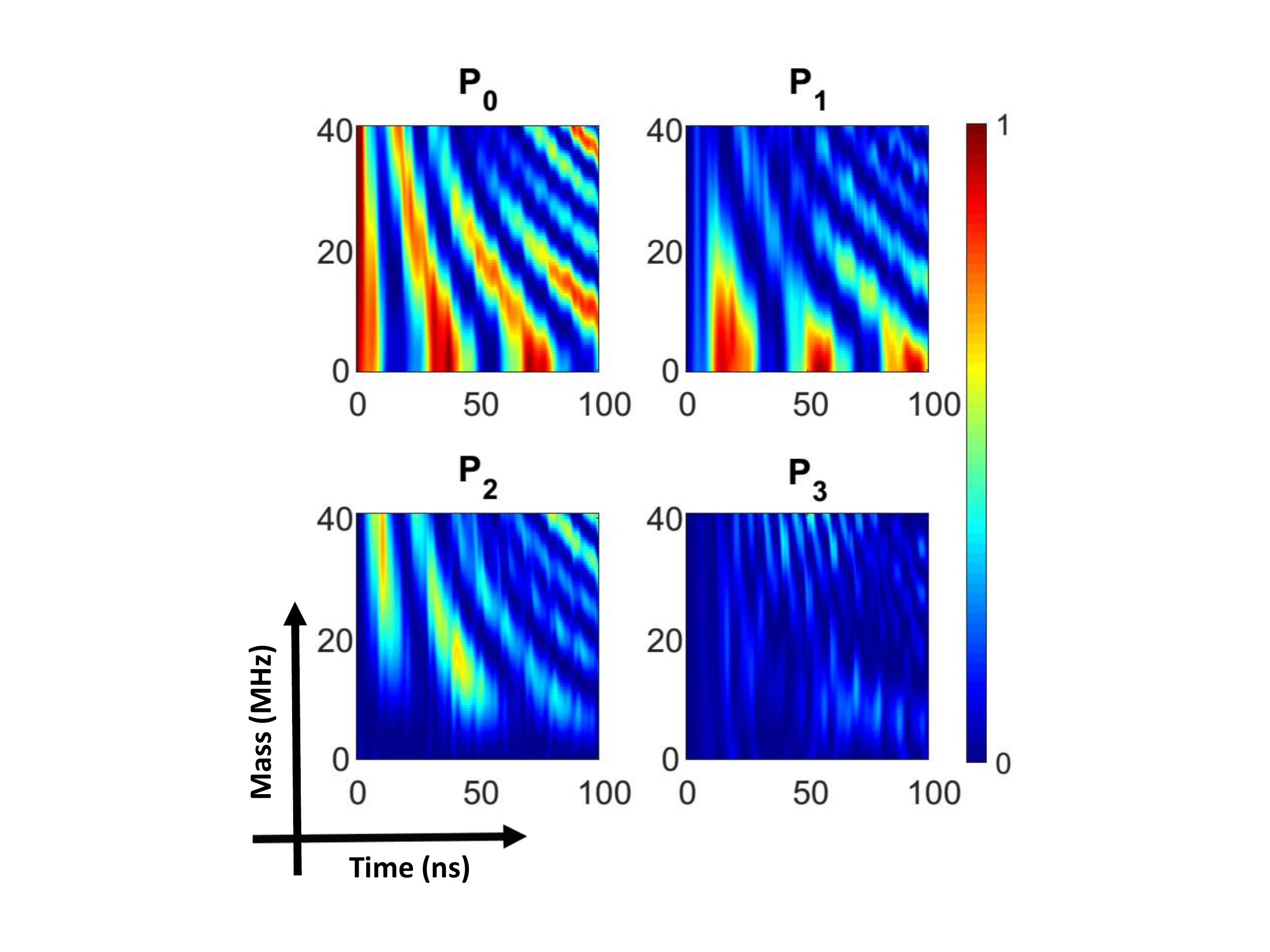} 
		\caption{Populations of the dressed diamond eigenstates constructed from Fig. S2. With the appropriate circuitry these may be measured directly.}
		\label{fig:S3}
	\end{figure*}

\end{document}